\documentclass[twocolumn,showpacs,preprintnumbers,prb]{revtex4}
\usepackage{amsmath}
\usepackage{graphicx}
\usepackage{dcolumn}
\usepackage{bm}
\usepackage{epsfig}

\begin{document}

\title{Magnetic field influence on the proximity effect in semiconductor -
superconductor hybrid structures and their thermal conductance}
\author{Grygoriy Tkachov$^{1,2}$ and Vladimir I. Fal'ko$^{1}$}
\affiliation{$^{1}$ Department of Physics, Lancaster University, Lancaster, LA1 4YB,
United Kingdom\\
$^{2}$ Institute for Radiophysics and Electronics NAS, Kharkiv, 61085,
Ukraine}
\date{\today }

\begin{abstract}
We show that a magnetic field can influnce the proximity effect in NS
junctions via diamagnetic screening current flowing in the superconductor.
Using ballistic quasi-one-dimensional (Q1D) electron channels as an example, we
show that the supercurrent flow shifts the proximity-induced minigap in the
excitation spectrum of a Q1D system from the Fermi level to higher
quasiparticle energies. Thermal conductance of a Q1D channel (normalized by
that of a normal Q1D ballistic system) is predicted to manifest such a
spectral feature as a nonmonotonic behavior at temperatures corresponding
to the energy of excitation into the gapful part of the spectrum.
\end{abstract}

\maketitle

The superconducting proximity effect is a mesoscopic scale phenomenon, which
consists in the penetration and coherent propagation of Cooper pairs from a
superconductor (S) into a normal metal (N). The Cooper pair transfer into the
normal metal can be equivalently described as an Andreev reflection process 
\cite{Andreev} which consists of electron (with momentum $\mathbf{p}$)
conversion into a Fermi sea hole (with momentum $-\mathbf{p}$) at the NS
interface. The interference between an electron and the Andreev reflected hole
imposes a minigap onto the spectrum of quasiparticle excitations near the
Fermi level in the normal part of such a hybrid structure \cite{Minigap},
thus giving rise to pronounced features in its I(V) characteristics 
\cite{vanWees,Petrashov,Pannetier} and thermoelectric properties \cite{Thermo}.
Studies of the proximity effect have recently been made in various combinations
of materials, including junctions between superconductors and semiconductor
structures \cite{vanWees} supporting a two-dimensional electron gas. In the
case of electrons in a semiconductor structure weakly coupled to a
superconductor, the minigap value discussed in the literature 
\cite{Volkov,MiniGExp} is much smaller than that of the 'mother' gap in the
superconductor, both due to the mismatch $v_{F}\ll v_{S}$ between Fermi
velocities in the two-dimensional gas [$v_{F}=(2E_{F}/m)^{1/2}$] 
and the superconducting metal ($v_{S}$), 
and also due to a possible Schottky barrier between them, with
transparency $\theta \sim e^{-2a/\lambda }$ (dependent on the length $\lambda $
of electron penetration into the barrier of the thickness $a$), 
$E_{g}\approx \frac{v_{F}}{v_{S}}\theta E_{F}\ll \Delta $.

It has been noticed that the electron-hole interferences and the SN
proximity effect survive at higher magnetic fields than the weak localisation -
another quantum interference effect~\cite{vanWees,Petrashov,Pannetier}. This
has been understood as a consequence of the fact that the interfering
electron and Andreev-reflected hole retrace the same geometrical path in the
normal metal, thus hardly encircling any magnetic flux~\cite{Cond-theory}.
Therefore, another mechanism of magnetic field influence on the superconducting
proximity needs to be taken into account, via a screening diamagnetic
supercurrent on the S-side of the hybrid structure. Since Andreev reflection
takes place at the NS interface, where Cooper pairs flow, the incoming
electron and the hole reflected by a moving condensate of Cooper pairs would
be slightly shifted in momentum space; hence the ideal condition for them
to retrace the same geometrical path is violated. As the orbital effect of
the magnetic field on the normal metal or semiconductor side of the system is
weak, the influence via diamagnetic screening may be the major factor of
magnetic field influence on the superconducting proximity effect.

Below, we analyze the influence of diamagnetic supercurrent in the system
where the latter would be the only way a magnetic field might affect the
proximity effect: a ballistic one-dimensional conductor connected in parallel
to a superconducting bulk [Fig.~1a]. To be specific, we model such a
conductor as a quasi-one-dimensional (Q1D) channel formed near the edge of a
2D electron gas in a heterostructure (x-y plane) with the side contact to a
superconducting film, by depleting the 2D gas using a split top gate, and
subjected to a weak magnetic field $\mathbf{B}=(0;0;B)$. We show that the
spectrum of low-energy quasiparticle excitiations in such a hybrid system
has the minigap displaced with respect to the Fermi level to higher energies,

\begin{equation}
\epsilon _{\alpha p}^{\pm }=v_{F}\Pi \times \mathrm{sgn}p-\alpha
\varepsilon _{Z}\pm \sqrt{v_{F}^{2}(|p|-p_{F})^{2}+E_{g}^{2}},
\label{Spectrum}
\end{equation}%
reflecting the fact that Cooper pairs in the channel are forced into the flow
while tunneling from the bulk of the superconductor (where they are formed
of two electrons with exactly opposite momenta) across the region of
penetration of the magnetic field. [The Zeeman splitting effect is also
taken care of by the term $\alpha \varepsilon _{Z}$ ($\alpha $ is the spin
projection) in Eq.~(\ref{Spectrum})]. As a result, each of the two
electrons acquires the momentum shift

\begin{equation}
\Pi =\frac{eB\delta }{c}\tanh \frac{L}{2\delta }  \label{Pi}
\end{equation}%
caused by the Lorentz force and equal to the difference between the vector
potential $\mathbf{A}=(0,A,0)$ deep inside the superconductor, $A=0$, and at
its surface, $A=B\delta \tanh (L/2\delta )$, where $\delta $ and $L$ stand
for the London penetration depth and the superconductor film thickness,
respectively. The spectrum described by Eq. (\ref{Spectrum}) can also be
understood as one of the Bogolubov quasiparticles in the laboratory frame,
where the equilibrium conditions are set by the heat reservoirs, for the
condensate moving along the Q1D channel with the drift velocity $\Pi /m$ ($m$
is the effective electron in the semiconductor). According to 
Eq.~(\ref{Spectrum}) the minigap is removed from the Fermi level when the field
reaches the value

\begin{equation}
B^{\ast }\approx B_{c1}\frac{\delta }{\xi }\coth \left( \frac{L}{2\delta }%
\right) \frac{\theta E_{F}}{\Delta }\ll B_{c1},  \label{B*}
\end{equation}%
where $B_{c1}$ and $\xi $ are the first critical field and the coherence
length in the superconductor.

The removal of a minigap from the Fermi level caused by a magnetic field would
manifest itself in the transport properties of a hybrid system, such as the
electron-mediated heat transfer. The ballistic quasiparticle spectrum in
Eq.~(\ref{Spectrum}) gives rise to the thermal conductance

\begin{equation}
\kappa (T,B)=\kappa _{N}(T)\times \frac{3}{4\pi ^{2}}\sum_{\ \pm
}\int\limits_{\frac{E_{g}\pm v_{F}\Pi }{k_{B}T}}^{\infty }\frac{~x^{2}dx}{%
\cosh ^{2}\frac{x}{2}},  \label{kappa}
\end{equation}%
where $\kappa _{N}(T)=\pi k_{B}^{2}T/3\hbar $ is the conductance of a normal
quantum ballistic wire~\cite{Pendry}. At a zero magnetic field, the
temperature dependence of $\kappa $ is activational, $\kappa
(T<E_{g}/k_{B})\propto e^{-E_{g}/k_{B}T}$, whereas at high fields, when
there is no gap at the Fermi energy, $\kappa (T,B)=\kappa _{N}(T)$. The
crossover from low to high fields takes place at $B^{\ast}$ [Eq.~(\ref{B*})] 
and reflects the presence of a minigap $E_{g}$ in the quasiparticle spectrum at
finite excitation energies. This results in a nonmonotonous temperature and
magnetic field dependence of the ratio $\kappa (T,B)/\kappa _{N}(T)$.

The analysis of the quasiparticle spectrum formed due to multiple Andreev
reflections in this paper is based on the standard weak-coupling approach to
the proximity effect description in superconductor junctions with normal
metals and electron layers in semiconductors~\cite{Volkov}. To be specific,
we describe the Q1D confinement (provided by a gate) by the 2D electron wave
function $\varphi (x)$ localized in the $x$ direction, whose magnitude at
the interface can be estimated from the boundary condition $\varphi
(0)=\lambda \partial _{x}\varphi (0)$, with $\lambda $ standing for the
electron penetration length into the barrier. The Fermi momentum of the Q1D
system $p_{F}$ and 3D electron density on the semiconductor side are assumed
to be much smaller than those in the superconductor, and we also take the
tunneling coefficient $\theta \sim \exp (-2a/\lambda )$ as a small parameter.
These assumptions enable us to neglect the influence of the normal system on
the superconductor and to investigate the proximity effect in the Q1D system
without feedback.

\begin{figure}[t]
%\vspace{0.3cm}
\hspace{0.15\hsize}
\epsfxsize=0.7\hsize \epsffile{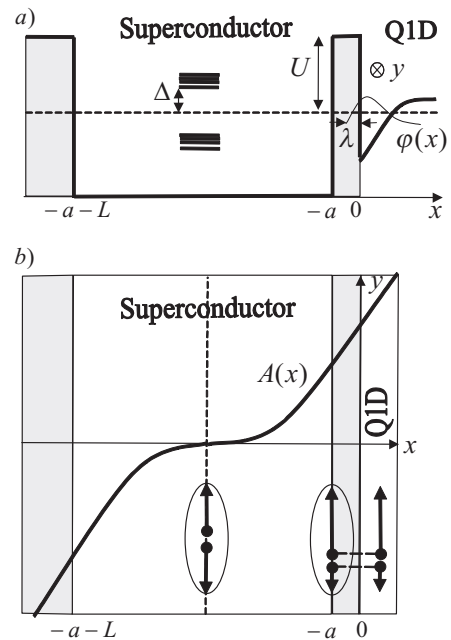} %%\vspace{0.0cm}
\caption{(a) Schematic view of a superconductor/Q1D system junction. 
(b) Vector potential profile.}
\label{fig:1}
\end{figure}

In the presence of a magnetic field $\mathbf{B}=(0;0;B)$ it is convenient to
choose the vector potential to be parallel to the interface, $\mathbf{A}%
(x)=[0,A(x),0]$ in order to deal with a real order parameter in the
superconductor. The vector potential $A(x)$ acting on the normal electrons
must be found self-consistently, taking into account the screening of the
external magnetic field $B$ by a diamagnetic supercurrent~\cite{Minigap,Zul}. 
Inside the superconductor $A(x)$ can be found from the London equation
with the boundary conditions $\partial _{x}A(-a)=B$ and 
$\partial_{x}A(-a-L)=B$ as follows:

\begin{equation*}
A(x)=B\delta \frac{\sinh [(x+a+L/2)/\delta ]}{\cosh (L/2\delta )}.
\end{equation*}%
It is antisymmetric with respect to the middle of the superconductor: 
$A(x=-a-L/2)=0$ [Fig.1(b)]. Since $A(x)$ must be continuous at the surface of
the superconductor $x=-a$, in the semiconductor $x\geq -a$ it varies as 
$A(x)=B(x+a)+B\delta \tanh (L/2\delta )$. The width of the electronic wave
function in the Q1D channel, $\delta x\sim k_{F}^{-1}$ and the barrier
thickness $a$ are both much less than $L$ or $\delta $, therefore, the
vector potential acting on the Q1D electrons is virtually a constant: 
$A(x)\approx A(-a)=B\delta \tanh (L/2\delta )$, which will be used below to
determine the quasiparticle spectrum in the channel.

We describe superconducting correlations in the Q1D channel using a pair of
coupled equations for $\hat{\psi}_{p}(t)=\left( 
\begin{array}{c}
\psi _{\alpha p}(t) \\ 
\psi _{-\alpha p}(t)%
\end{array}%
\right) $ \ and $\hat{\psi}_{p}^{\dag }(t)=\left( 
\begin{array}{c}
\psi _{\alpha p}^{\dag }(t) \\ 
\psi _{-\alpha p}^{\dag }(t)%
\end{array}%
\right) $ -- creation and annihilation operators:

\begin{eqnarray}
&&\left[ i\hbar \partial _{t}-\frac{(p+\Pi )^{2}}{2m}+\sigma _{3}\varepsilon
_{Z}+E_{F}\right] \hat{\psi}_{p}(t)=  \label{Eq1} \\
&=&\vartheta^{1/2}\int dt^{\prime }\left[ G(t,t^{\prime })\hat{\psi}%
_{p}(t^{\prime })+F^{\ast }(t,t^{\prime })i\sigma _{2}\hat{\psi}_{-p}^{\dag
}(t^{\prime })\right] ,  \notag
\end{eqnarray}
\begin{eqnarray*}
&&\left[ -i\hbar \partial _{t}-\frac{(-p+\Pi )^{2}}{2m}+\sigma
_{3}\varepsilon _{Z}+E_{F}\right] \hat{\psi}_{-p}^{\dag }(t)= \\
&=&\vartheta^{1/2}\int dt^{\prime }\left[ F(t,t^{\prime })i\sigma
_{2}^{t}\psi _{p}(t^{\prime })+G^{\ast }(t,t^{\prime })\hat{\psi}_{-p}^{\dag
}(t^{\prime })\right] ,
\end{eqnarray*}
where 
$\vartheta =\left(\frac{\hbar ^{2}\varphi (0)}{m_{B}\lambda \exp (a/\lambda )}
\right)^2\sim 
\left(\frac{\hbar ^{2}k_{F}^{3/2}}{m\exp (a/\lambda )}\right)^2$ 
characterizes the tunneling coupling to the superconductor and the electron momentum 
shift in the magnetic field $\Pi $ is related to the vector potential by Eq. (\ref{Pi}). 
In Eq. (\ref{Eq1}), $G(t,t^{\prime })\equiv G(x=-a,x^{\prime
}=-a,t-t^{\prime })$ and $F(t,t^{\prime })\equiv F(x=-a,x^{\prime
}=-a,t-t^{\prime })$ are the normal and anomalous Green functions of the
superconductor at its boundary; $\sigma _{2}$ and $\sigma _{3}$ are Pauli
matrices ($\sigma ^{t}$ is transposed to $\sigma $). Since the size of the
Fermi sea in the semiconductor wire is much smaller than in the
superconductor, one can ignore the dependence of $G$ and $F$ on the momentum
parallel to the interface: only electrons in the superconductor moving
nearly perpendicularly to the interface can tunnel into the Q1D wire. Since we
are interested in the low-temperature regime $k_{B}T\sim E_{g}\ll \Delta $,
we will neglect the terms containing the normal Green function $G$ in 
Eqs.~(\ref{Eq1}). For the chosen gauge, the anomalous Green function of the
superconductor, $F$ in Eqs.~(\ref{Eq1}) has no phase factors, despite the
presence of a magnetic field. For a weak field $B\ll B_{c1}$, its time
Fourier transform can be estimated as $F(\epsilon )\approx
L^{-1}\sum\nolimits_{p_{x}}\Delta /(\Delta ^{2}-\epsilon ^{2}+\eta
_{p_{x}}^{2})$, with $\eta _{p_{x}}$ being the normal electron dispersion
near the Fermi level in the superconductor. The integration over the
perpendicular momentum $p_{x}$ gives 
$F(\epsilon )\approx \Delta /\hbar v_{S}(\Delta ^{2}-\epsilon ^{2})^{1/2}$, 
thus giving us the minigap $E_{g}=\vartheta F(\epsilon =0)$ 
mentioned in the introduction and obtained in earlier publications~\cite{Volkov}.

The solution of Eqs. (\ref{Eq1}) for $\epsilon \ll \Delta $ is given by the
Bogolubov transformation of the form

\begin{eqnarray}
\psi _{\alpha p}(t) &=&u_{p}b_{\alpha p}\exp (-it\epsilon _{\alpha
p}^{+}/\hbar )  \label{sol} \\
&&+i\sigma _{2}^{\alpha ,-\alpha }v_{p}b_{-\alpha -p}^{\dag }\exp
(-it\epsilon _{\alpha p}^{-}/\hbar )  \notag \\
u_{p}^{2} &=&\frac{1}{2}\left[ 1+\frac{v_{F}(|p|-p_{F})}{%
[v_{F}^{2}(|p|-p_{F})^{2}+E_{g}^{2}]^{1/2}}\right]  \notag \\
v_{p}^{2} &=&1-u_{p}^{2}  \notag
\end{eqnarray}%
where $b_{\alpha p}$ and $b_{-\alpha -p}^{\dag }$ are Bogolubov's
quasiparticle operators, and the excitation spectrum $\epsilon _{\alpha
p}^{\pm }$ is given by Eq. (\ref{Spectrum}) (see Fig.~2). The Zeeman term
in Eq.~(\ref{Spectrum}) turns out to be much smaller than the orbital one: 
$E_{Z}/v_{F}\Pi \sim g/k_{F}\min (\delta ,L)\ll 1$ -- unless the electron
g-factor is anomalously large.

Due to the motion of the Q1D condensate the excitation energy curve is
tilted by energy $v_{F}\Pi \,\mathrm{sign}p$. The field $B^{\ast }$ [Eq.~(\ref{B*})] 
at which the minigap is removed from the Fermi level is determined by the
condition that $v_{F}\Pi =E_{g}$. Note that at higher fields $B^{\ast }<B\ll
B_{c1}$, the quasiparticle spectrum remains gapful, 
with the center of the
gap moved to energies $\sim E_{g}$.

\begin{figure}[t]
%\vspace{0.3cm}
\hspace{0.15\hsize}
\epsfxsize=0.7\hsize \epsffile{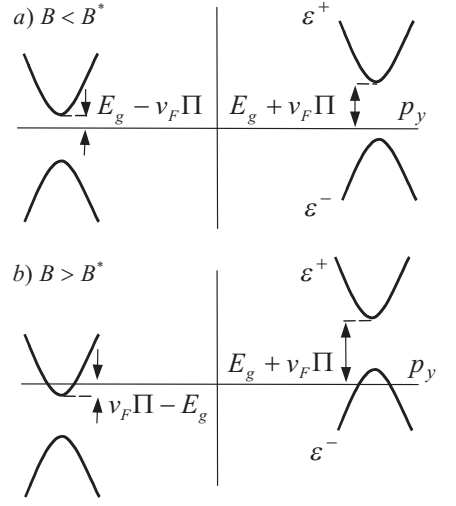} %\vspace{0.0cm}
\caption{Schematic view of the quasiparticle spectrum described by 
Eq.~(\protect\ref{Spectrum}).}
\label{fig:2}
\end{figure}

Now we turn to the calculation of the thermal conductance $\kappa (T,B)$ of
a long Q1D channel whose ends are kept at temperatures $T$ and $T+\Delta T$ $%
(\Delta T\ll T)$. Since no heat can get into the strongly gaped
superconductor, the middle of the wire represents a bottle-neck for the
heat transport, so that we can analyze $\kappa (T,B)$ in the infinite wire
geometry. The expression for the energy current operator $j_{\epsilon }(yt)$
in a wire can be found from the continuity equation $\partial
_{y}j_{\epsilon }(yt)=-\partial _{t}\rho _{\epsilon }(yt)$, where the density
of energy $\rho _{\epsilon }(yt)$ corresponding to the equations of motion (%
\ref{Eq1}) is

\begin{gather}
\rho _{\epsilon }(yt)=\frac{1}{2}\sum\limits_{\alpha }\left( \psi _{\alpha
}^{\dag }(yt)\left[ \frac{(\hat{p}+\Pi )^{2}}{2m}-\alpha \varepsilon
_{Z}-E_{F}\right] \psi _{\alpha }(yt)\right.  \notag \\
\left. -E_{g}i\sigma _{2}^{\alpha ,-\alpha }\psi _{-\alpha }^{\dag }(yt)\psi
_{\alpha }^{\dag }(yt)+H.c.\right) ,
\end{gather}%
where $\psi _{\alpha }(yt)=L_{y}^{-1/2}\sum_{p}\psi _{\alpha }(t)\exp
(ipy/\hbar )$ with $L_{y}$ being the length of the Q1D system, and $\hat{p}%
=-i\hbar \partial _{y}$. Using the Bogolubov transformation (\ref{sol}) for $%
\psi _{\alpha }(t)$, for the density of energy one finds

\begin{eqnarray}
&& \rho_{\epsilon}(yt)=\sum\limits_{\alpha pp^\prime} \frac{e^{\frac{iy}{%
\hbar}(p^\prime -p)}}{2L_{y}}\times  \label{rho} \\
&& \times \{v_{p}v_{p^\prime}\ (\epsilon_{\alpha
p^\prime}^{-}+\epsilon_{\alpha p}^{-}) b_{-\alpha -p}b_{-\alpha
-p^\prime}^{\dagger}\ e^{-\frac{it}{\hbar}(\epsilon_{\alpha
p^\prime}^{-}-\epsilon _{\alpha p}^{-})}+  \notag \\
&& +u_{p}u_{p^\prime}\ (\epsilon_{\alpha p^\prime}^{+}+\epsilon_{\alpha
p}^{+}) b_{\alpha p}^{\dagger}b_{\alpha p^\prime}\ e^{-\frac{it}{\hbar}%
(\epsilon_{\alpha p^\prime}^{+}-\epsilon_{\alpha p}^{+})}+  \notag \\
&& +u_{p}v_{p^\prime}\ i\sigma _{2}^{\alpha ,-\alpha } (\epsilon_{\alpha
p}^{+}+\epsilon_{\alpha p^\prime}^{-}) b_{\alpha p}^{\dagger }b_{-\alpha
-p^\prime}^{\dagger } e^{\frac{it}{\hbar}(\epsilon_{\alpha
p}^{+}-\epsilon_{\alpha p^\prime}^{-})}+  \notag \\
&& +v_{p}u_{p^\prime}\ i\sigma _{2}^{-\alpha ,\alpha } (\epsilon_{\alpha
p}^{-}+\epsilon_{\alpha p^\prime}^{+}) b_{-\alpha -p}b_{\alpha p^\prime} e^{%
\frac{it}{\hbar}(\epsilon_{\alpha p}^{-}-\epsilon_{\alpha p^\prime}^{+})} \}.
\notag
\end{eqnarray}
In order to satisfy the continuity equation with $\rho_{\epsilon}(yt)$ given
by Eq. (\ref{rho}) the energy current $j_{\epsilon }(yt)$ must have the
following form:

\begin{eqnarray}
&& j_{\epsilon}(yt)=\sum\limits_{\alpha pp^\prime} \frac{e^{\frac{iy}{\hbar}%
(p^\prime -p)}}{2L_{y}}\times  \label{jop} \\
&& \times \{v_{p}v_{p^\prime}\ \frac{(\epsilon_{\alpha
p^\prime}^{-})^{2}-(\epsilon_{\alpha p}^{-})^{2}} {p^\prime-p}b_{-\alpha
-p}b_{-\alpha -p^\prime}^{\dagger}\ e^{-\frac{it}{\hbar}(\epsilon_{\alpha
p^\prime}^{-}-\epsilon _{\alpha p}^{-})}+  \notag \\
&& +u_{p}u_{p^\prime}\ \frac{(\epsilon_{\alpha
p^\prime}^{+})^{2}-(\epsilon_{\alpha p}^{+})^{2}} {p^\prime-p}b_{\alpha
p}^{\dagger}b_{\alpha p^\prime}\ e^{-\frac{it}{\hbar}(\epsilon_{\alpha
p^\prime}^{+}-\epsilon_{\alpha p}^{+})}\}.  \notag
\end{eqnarray}
In Eq. (\ref{jop}) we have already omitted the terms containing $b_{\alpha
p}^{\dagger }b_{-\alpha -p^\prime}^{\dagger }$ and $b_{-\alpha -p}b_{\alpha
p^\prime}$ which vanish after the averaging. The averaged value of the
energy current $j_{\epsilon}$ can be written as the sum of two contributions:

\begin{equation}
j_{\epsilon }=-h^{-1}\sum_{\alpha }\int dp\ \epsilon _{\alpha
p}^{+}\partial_{p} \epsilon _{\alpha p}^{+}v_{p}^{2}+j_{q}.  \label{je}
\end{equation}
The first of them can be attributed to the supercurrent flow and cannot
transfer heat, whereas $j_{q}$ represents the heat current:

\begin{equation}
j_{q}=h^{-1}\sum_{\alpha }\int dp\ \epsilon _{\alpha p}^{+}\partial
_{p}\epsilon _{\alpha p}^{+}n(\epsilon _{\alpha p}^{+}).  \label{jq}
\end{equation}%
The latter is determined by the energy distributions, $n(\epsilon _{\alpha
p}^{+})$ and the group velocity, $\partial _{p}\epsilon _{\alpha p}^{+}$ of
quasiparticles. We express the energy currents (\ref{je}) and (\ref{jq}) in
terms of the "$+$"-branch of the spectrum (\ref{Spectrum}) using the
relationship $\epsilon _{\alpha p}^{-}=-\epsilon _{-\alpha -p}^{+}$ and the
symmetry of the limits in the sum. The distribution functions of rightmovers
($\partial _{p}\epsilon _{\alpha p6}^{+}>0$) and leftmovers ($\partial
_{p}\epsilon _{\alpha p}^{+}<0$) are assumed to be different and set by
reservoirs, as $n(\epsilon _{\alpha p}^{+},T+\Delta T)$ and $n(\epsilon
_{\alpha p}^{+},T)$, respectively. Using this, we determine the thermal
conductance $\kappa (T,B)$ given by Eq. (\ref{kappa}) as the proportionality
coefficient between the heat current and the temperature drop, $j_{q}=\kappa
(T,B)\Delta T$.

\begin{figure}[!t]
%\vspace{0.3cm}
\hspace{0.08\hsize}
\epsfxsize=0.7\hsize \epsffile{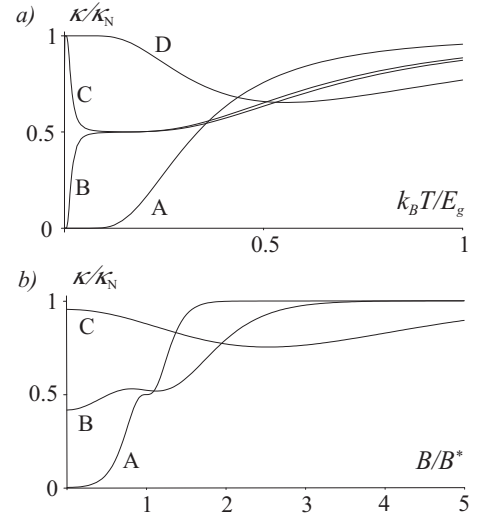} %\vspace{0.1cm}
\caption{(a) Temperature dependence of the thermal conductance, $\protect%
\kappa$ normalized by that of a normal wire, $\protect\kappa_{N}$ for
different values of magnetic field: (A) $B/B^*=0.01$, (B) $B/B^*=0.95$, 
(C) $B/B^*=1.05$, and (D) $B/B^*=2$. (b) Magnetic field dependence for different
temperatures: (A) $k_BT/E_g=0.1$, (B) $k_BT/E_g=0.3$, and (C) $k_BT/E_g=1$.}
\label{fig:3}
\end{figure}

Figure 3(a) shows the thermal conductance (\ref{kappa}) normalized by that of
a normal wire as a function of $k_BT/E_{g}$ for different values of the
magnetic field. Plot A is related to $B=0$ and shows how the conductance
exponentially decreases at temperatures smaller than the minigap $E_{g}$.
Curves B and C show what happens when the field crosses the value of $%
B^{\ast }$, at which the edge of the minigap is about to reach the Fermi
level. For $B<B^{\ast}$ (curve B), $\kappa (T)/\kappa _{N}(T)$ is
exponentially small only if $k_BT<E_{g}-v_{F}\Pi \ll E_{g}$. When the
temperature is in the interval $E_{g}-v_{F}\Pi <k_BT<E_{g}+v_{F}\Pi $,
quasiparticles with negative momenta $p\approx -p_{F}$ transfer heat,
whereas the states with positive $p$ are still unpopulated. This interval
corresponds to the plato in curve B where the conductance $\kappa (T)$ is
half of that in the normal state. At higher temperatures, $%
k_BT>E_{g}+v_{F}\Pi $ the asymmetry of the excitation spectrum no longer
matters, and $\kappa (T)\approx \kappa _{N}(T)$.

When the field exceeds $B^{\ast }$ (curves C and D), the dependence $\kappa
(T)/\kappa _{N}(T)$ becomes nonmonotonic. As in a normal wire, at low
temperatures $k_{B}T\ll v_{F}\Pi -E_{g}$ there are two left-moving and two
right-moving modes capable of tranferring heat, which gives $\kappa
(T)=\kappa _{N}(T)$. At intermediate temperatures $v_{F}\Pi -E_{g}\ll
k_{B}T\ll E_{g}+v_{F}\Pi $, only the states with negative momenta contribute
to the thermal conductance: $\kappa (T)=\kappa _{N}(T)/2$. At higher
temperatures the conductance recovers a normal metallic behaviour. Finally,
when $B\gg B^{\ast }$ the minimum in $\kappa (T)/\kappa _{N}(T)$ is less
pronounced and the heat conductance behaviour becomes indistinguishable from
that of a normal wire. The magnetic field dependence of $\kappa /\kappa _{N}$
is given in Fig. 3(b).

The authors thank U. Zulicke, I. Aleiner, and A. Geim for useful discussions. 
This work was funded in parts by EPSRC (UK) and EC STREP within the
Framework 6 EU programme.

%%%%%%%%%%%%%%%%%%%%%%%%%%%%%%%%%%%%%%%%%%%%%%%%%%%%%%%%%%%%%%%%%%%%%%%%%%%%%%%%%%%%%%%%%%%%%%%%%

\newpage

\end{document}